# Kinematic structure of the Milky Way galaxy,
# near the spiral arm tangents

Jacques P Vallée

Herzberg Astronomy and Astrophysics, National Research Council of Canada

ADDRESS   5071 West Saanich Road, Victoria, British Columbia, Canada V9E 2E7

ORCID  *http://orcid.org/0000-0002-4833-4160*

EMAIL   jacques.p.vallee@gmail.com

KEYWORDS  astrophysics   -   Galaxy    -  Milky Way     -  spiral arms    -   symmetries

**Abstract.**

We compare the observed radial velocity of different arm tracers, taken near the tangent to a spiral arm. A slight difference is predicted by the density wave theory, given the shock predicted at the entrance to the inner spiral arm.  In many of these spiral arms, the observed velocity offset confirms the prediction of the density wave theory (with a separation between the maser velocity and the CO gas peak velocity, of about 20 km/s) - when the observed offset is bigger than the error estimates.

1.   **Introduction.**

How to get the precise location of a spiral arm? We could draw a spiral fit through parallax-distance data of interferometric-based radio masers or optical Gaia DR3 young stars in spiral arms, and a fit to an arm model. Or we could find the tangents from the sun to tracers inside a spiral arm (galactic longitudes), and a fit to an arm model. Both approaches give the same locations for spiral arms. Here we will use this second approach.

The arm tangent to a spiral arm is a line from the Sun to that spiral arm, being tangent to the arm (not crossing the arm). It should be mentioned that the tangent from the Sun to a spiral arm, done several times in different arm tracers, provides a **precise** galactic longitude on which to fit an arm model. Such a catalog of over 200 observed arm tangents has been published [1],[2],[3].  It was found observationally that each arm tracer was offset from each other arm tracer [4]: radio masers near the inner arm edge, but broad diffuse CO gas peaking at the outer arm edge, and many other tracers peaking in between.

For stars and gas in Galactic quadrants I and IV, one can look tangentially to a spiral arm. Using one arm tracer, a telescope scan shows a consistent value in Galactic longitude, from one telescope to the next.

While making a telescope drift in galactic longitude, along the disk of the Milky Way galaxy, we can record the intensity of an arm in a tracer (maser, HII regions, broad diffuse CO gas peaks, etc).  When the telescope sweeps across  a spiral arm width, there will be a galactic longitude where  the intensity of that arm tracer increases, peaks, and then decreases; so we will record the galactic longitude where the peak intensity was located in that arm tracer. If possible, the observers also recorded the radial velocity as observed at that peak.

We published a 4-arm spiral model as fitted to the tangent in broad diffuse CO gas peaking in each spiral arm (see [5],[6],  for the basic equations). A later fit [7] was done, with more data and with improved Galactic parameters: 8.15 kpc for the Sun's distance to the Galactic Center [8]. Other arm parameters are the arm pitch angle = 13.1$^O$, and the arms start at 2.2 kpc from the Galactic Center. Fitting uncertainties have been explained in [7]. The start of the Norma spiral arm in Galactic quadrant I is thus at $r_o$= 2.2 kpc and at an angle -30$^o$ below the horizontal line at the Galactic Center (perpendicular to the sun-to-Galactic Center line-of-sight)..

This global arm pitch angle was found earlier using a fit of arm segments well over both Galactic quadrants I and IV, enabling  better precision (Table 1 in [9]; Tables 1 and 2 in [10];  Fig. 4 in [7]);  small localised pitch deviations along the Galactic radius are thus smoothed out (see Fig. 1 in [11]).  Velocity wise, we took 233 km/s for the circular orbital velocity of the Local Standard of Rest around the Galactic Center [12].

For a more complete overview of the arm tangents as related to the Milky Way disk structure, see [6].

From published arm tracer separations and ages, the relative speed of the gas away from the arm shock front is estimated near 81 km/s –see [13]; [1]; [7]. In addition, a superposition of the known Galactic magnetic field can be made over the model spiral arm above, indicating that a counter-clockwise magnetic field covers the Sagittarius arm in Galactic Quadrant I and the Crux-Centaurus arm in Galactic Quadrant IV; the other arm segments and all other arms have a clockwise Galactic magnetic field [14].

In this paper, we check the predictions of the density wave theory regarding the speed of some tracers, relative to each another tracer. Section 2 deals with different radial velocities for different arm tracers. Section 3 deals with Galactic dynamics and the density wave theory. Section 4 compares the locations of both approaches (parallax, arm tangent). Section 5 shows a concluding discussion.

2. **New results.**

**Kinematic velocity of the galactic disk**. At arm tangent points, one can observe the radial velocity of specific arm tracers, such as the broad diffuse CO gas or of the radio masers. Being taken at separate galactic longitudes, their velocities must differ somewhat. Having gone through a shock front at different times in the density wave, they must respond differently.

We adopt the version of the density wave theory with shocks, in which the gas flow enters the arm at a supersonic velocity and creates a shock, and later the gas leaves the arm at a **subsonic** velocity (see fig. 3 in [15]). Going from one arm to the next, the gas orbit looks like a pointed oval streamline with a sharp bend at each shock location (see fig. 3 in [16]. The orbit is thus not quite circular around the Galactic Center, as typical excursions in azimuthal and radial velocities are about 20 km/s (Fig. 12 and Fig.13 in [16].

**Table 1** assembles the line of sight radial velocity values, as observed in some tangents to spiral arms in the Milky Way galaxy [1]. Also, mean results are shown in Table 1 and **Figure 2.** The label C is for the broad diffuse CO gas near the Potential Minimum of the density wave (and other blue tracers nearby, on the outer arm side). The label M is for the Maser data located near the shock of the density wave (and other orange tracers nearby, on the inner arm side). Both means for label C and label M are always within 20 km/s of each other. Typical errors bars are ±5 km/s.

**Table 1 – Mean radial velocity of each arm tracer, at each arm tangent [a]**

| | | | | | | | | |
|---|---|---|---|---|---|---|---|---|
| Mean tangent Longit.: | 283° | 310° | 328° | 338° | 346° | 018° | 030° | 050° |
| At Gal. radius (kpc): | 8.0 | 6.3 | 4.5 | 3.2 | 2.5 | 2.8 | 4.2 | 6.3 |
| Chemical Tracer: | $V_{rad}$ in Carina arm | $V_{rad}$ in Crux-Centaurus arm | $V_{rad}$ in Norma arm | $V_{rad}$ in Start of Perseus arm | $V_{rad}$ in Start of Sagittarius arm | $V_{rad}$ in Start of Norma arm | $V_{rad}$ in Scutum arm | $V_{rad}$ in Sagittarius arm |
| | (km/s) | (km/s) | (km/s) | (km/s) | (km/s) | (km/s) | (km/s) | (km/s) |
| **blue group:** | | | | | | | | |
| $^{12}CO$ at 8' | -8.8 | -46.6 | -97.6 | -126.7 | -136 | +125 | +95.0 | +55.3 |
| [CII] at 80" | - | - | -106[d] | -120[e] | - | - | - | - |
| [CII] at 12" | - | - | - | - | - | - | +114[b] | - |
| HI atom | -9 | -44 | -79 | - | - | - | - | - |
| HII complex | - | - | - | - | - | - | +100.0 | +61 |
| $^{13}CO$ | - | -35 | -85 | -115 | - | - | +95.0 | +60 |
| **Blue mean radial vel.:** | **-9±5** | **-42±5** | **-92±5** | **-121±5** | **-136±5** | **+125±5** | **+101±5** | **+59±5** |
| **orange group:** | | | | | | | | |
| [CII] at 80" | - | - | -99[f] | -127[g] | - | - | - | - |
| [Cii] at 12" | - | - | - | - | - | - | +115[c] | - |
| Warm $^{12}CO$ cores | - | - | - | - | - | - | +95 | +60 |
| Masers | +10 | -56.5 | -102 | -106.7 | -120 | +105 | +100.8 | +65.5 |
| **Orange mean radial vel.:** | **+10±5** | **-56±5** | **-101±5** | **-117±5** | **-120±5** | **+105±5** | **+104±5** | **+63±5** |
| Orange - Blue radial vel.: | +19±5 | -14±5 | -9±5 | +4±5 | +16±5 | -20±5 | +3±5 | +4±5 |
| Masers – $^{12}CO$ radial vel | +19±5 | -10±5 | -4±5 | +20±5 | +16±5 | -20±5 | +6±5 | +10±5 |

Notes:
(a): Some data are referenced. All other data from Table 1 in [1],
(b): Table 1 in Velusamy et al [17] for longitudes l=28° to 30°.
(c): Table 1 in Velusamy et al [17] for longitudes l=31°-33°.
(d): Fig.7b in Velusamy et al [18] – ll= 327° – 329°.
(e): Fig.8c in Velusamy et al [18] – l= 334°–336°.
(f): Fig.7a in Velusamy et al [18] – l=329°-331°.
(g): Fig.8b in Velusamy et al [18] – l=336°-338°.

**Figure 1** shows the results. Arm tangents (a line from the Sun tangentially to the arm) to the spiral arms are observed in Galactic quadrant I at the Sagittarius arm (near l=050º), at the Scutum arm (near l=030º), at the Norma arm (near l=18º ), and in Galactic quadrant IV at the Carina arm (near l= 283º), at the Crux-Centaurus arm (near l=310º), at the Norma arm (near l=328º), at the Perseus start arm (near l=338º), and at the Sagittarius start arm (near l=346º).

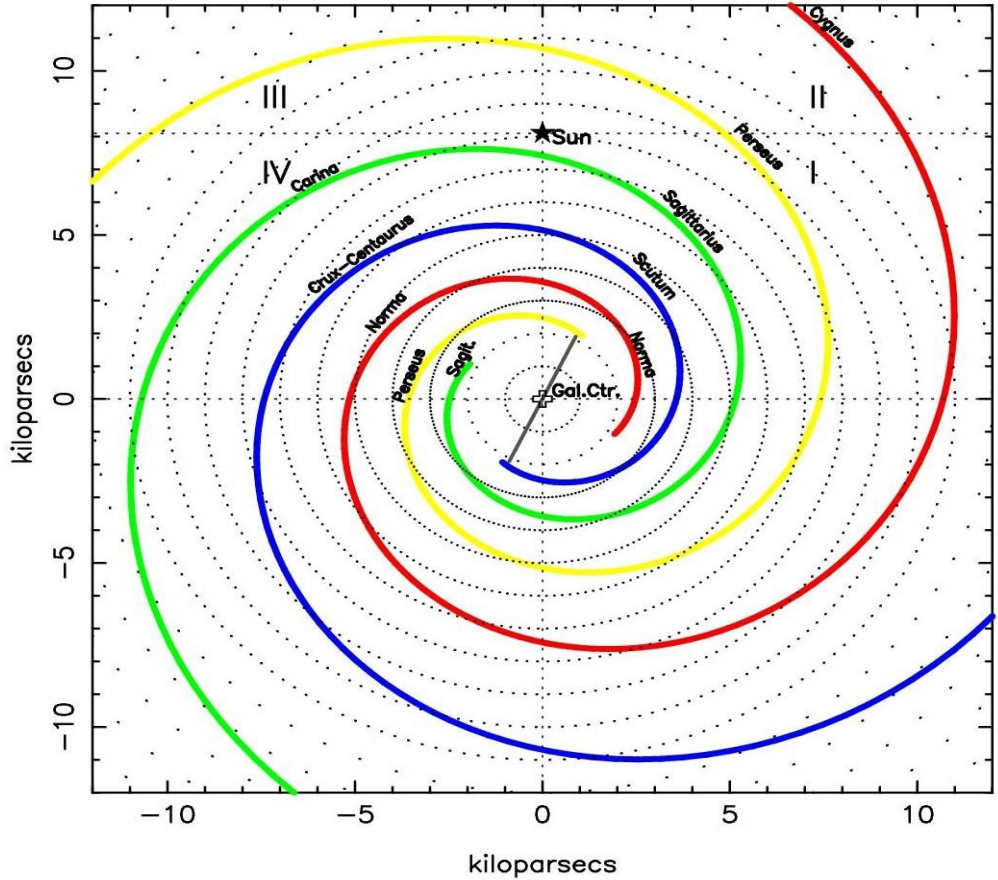

*Figure 1. The view of the Galactic disk, seen from above. Galactic quadrants I, II, III, and IV are indicated. The Sun is indicated at 8.15 kpc from the Galactic Center. Four spiral arms are shown, each in a different color. In this rendering, the arm pitch angle = 13.1º, and each arm starts at 2.2 kpc from the Galactic Center.*

**Figure 2** shows the results. In Galactic quadrant I, where all tracers should have a positive velocity, the slower M label should have a slightly smaller value than the faster C label, giving a negative value in the last row of Table 1. The M label is close to the shock front in the inner arm side, having a general slowing down

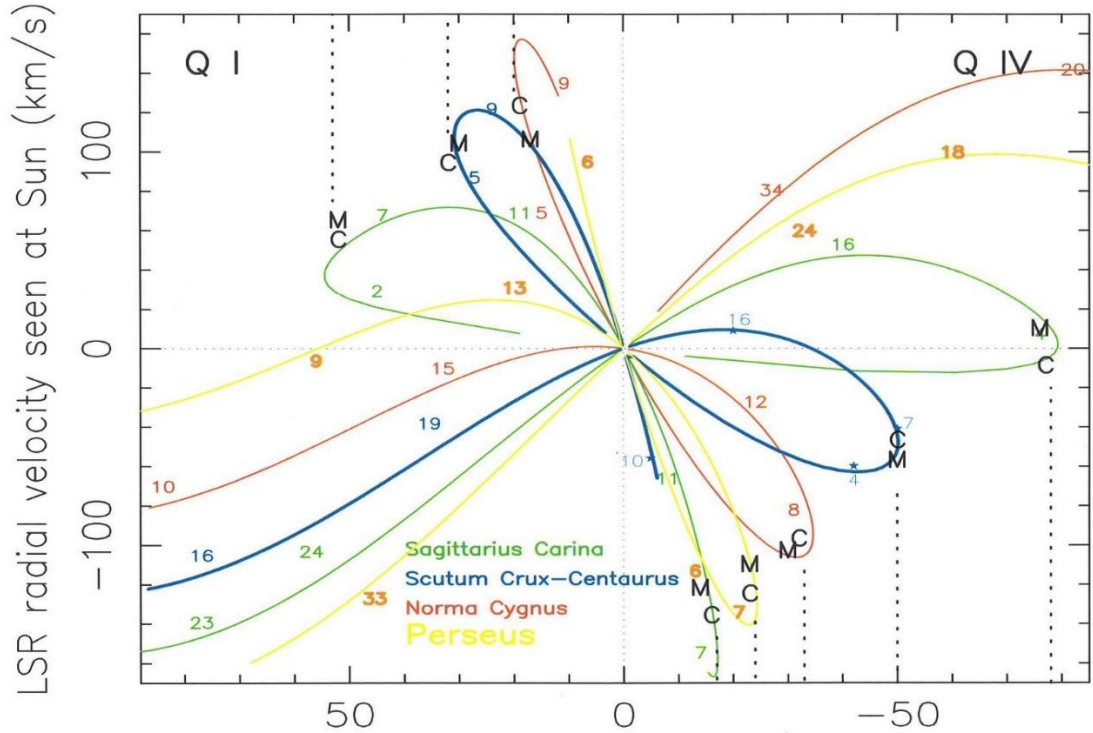

*Figure 2. Radial velocity of each spiral arm. The actual speed (vertical axis; in km/s) is given at each Galactic longitude (horizontal axis, in degrees), while the number shown near each arm is the actual distance (kpc) from the Sun. The arm tangents are shown by vertical dashed lines (from the bottom or the top). The actual location (radial velocity, Galactic longitude) of the radio masers (M) or the broad diffuse CO (C) at the arm tangent can be seen. The orbital circular speed used around the Galactic Center = 233 km/sec. As before, the arm pitch angle = 13.1°, while the arm starts at 2.2 kpc from the Galactic Center, and the Sun is distanced from the Galactic Center by 8.15 kpc.*

**[CII] gas.** In addition to masers and diffuse CO gas, we can add the [CII] line observed at a wavelength of 158 microns: Velusamy et al [17] at 12" resolution around the Scutum spiral arm near l= 30° ; Velusamy et al [18] at 80" resolution around the Norma arm near l= 328° and the Perseus arm near l= 338°.   In these [CII] data, both the on-tangent (near the maser tangent and the shock) and an off-tangent (near the diffuse CO tangent and the outer arm side) were measured.  These data are also reported in Table 1.

The [CII] at 80'' are taken at different galactic longitudes, thus some of which being 'on-tangent' and some being 'off-tangent' in their observations - see notes at bottom of table (matching the longitudes of other arm tracers).

**Prediction:** The density wave theory would predict in Galactic Quadrant IV that the orbiting CO gas  (negative speed) would win over the slower maser (M) speed (as observed in Table 1 for the Carina arm at 283°, the Perseus start arm at 338°, the Sagittarius start arm at 346°), and predict in Galactic quadrant I that the orbiting CO gas (positive speed)  would win over the slower maser speed (as observed in Table 1 for the Norma arm at 018°).   Consequently, requiring all arms with an offset larger than 3 times the velocity error, then all 4 remaining arms satisfy the prediction of the density wave  theory.

It does not seem the case for 4 arms:  the Sagittarius arm at l=50°, the Scutum arm at =30°,  the Norma arm at 328°, the Crux-Centaurus arm at 310°;  for these 4 arms, the velocity offset is smaller than 3 times the velocity error, so the offset is not significant.

3.  **Galactic dynamics**.

It has been observed that each arm tracer is separated from other arm tracers (see [4], [3]), as the orbiting gas flows through a spiral arm, entering at a shock in the inner arm side, then forming protostars, masers, proto-HII regions, and exiting on the outer arm side near the Potential Minimum of the density wave [16].

Recent measured values have been made for the distance of the Sun to the Galactic Center (8.15 kpc; [8]) and the mean speed of the Local Center of Rest near the Sun (233 km/s; [12]).

**Observed speed for new stars to flee the shock front/inner arm edge**. A recent compilation of arm tangents, using many arm tracers, was provided by [1]; that paper computed an age gradient of 11.3 ± 2 Myr/kpc, or a relative speed of arm tracers of 87 ± 10 km/s away from the shock front (inner arm edge). A very similar result from two different methods gave 12.0 ± 2 Myr/kpc and 81 ± 10 km/s [13], and 12.9 ± 2 Myr/kpc and 76 ± 10 km/s [7]. Taking here a statistical mean value would then give 12.1 ± 1 Myr/kpc and 81.3 ± 5 km/s.

This relative speed value would mostly apply at a Galactic radius near 7 kpc, where most masers are located - near the Sagittarius arm in Galactic quadrant I (see masers in Fig. 1 in [19]).

At the masers' orbit **near 7 kpc of galactic radius**, the linear density wave's pattern speed (shock) would be 233 – 81.3 = 151.7 km/s; thus the angular pattern speed at 7 kpc is 151.7 / 7.0 = 21.7 km/s/kpc. The co-rotation radius would then be 233/ 21.7 = 10.7 ±1 kpc, just beyond the Perseus arm along the Galactic Meridian (Sun to Galactic Center line). The co-rotation radius is where the gas going at orbital speed equals the linear value of the angular pattern speed.

**Figure 3** shows the speeds mentioned for the density wave's arm pattern, the new starforming masers, and the orbiting gas around the Galactic Center.

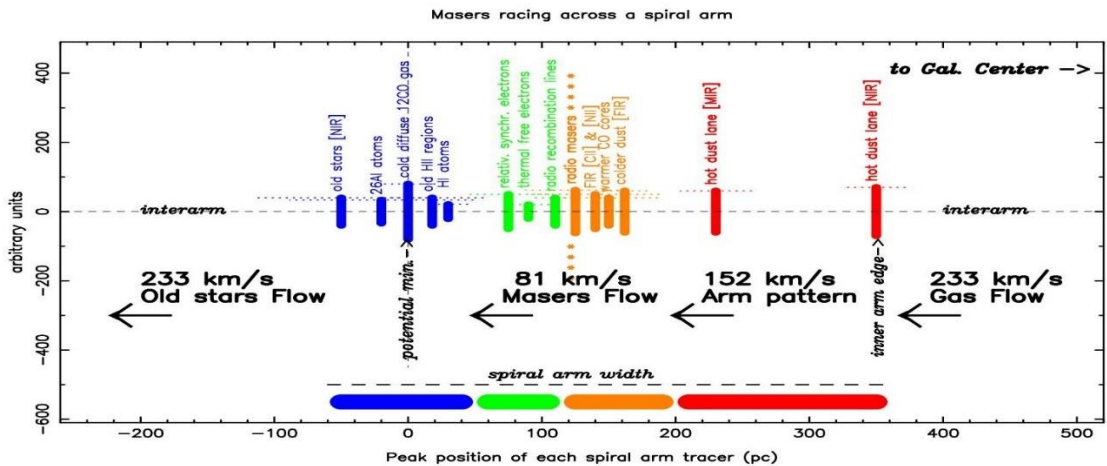

*Figure 3. A typical cross-section of a spiral arm, running over 350 parsecs from the inner arm edge (at right) to the Potential Minimum (at left). Each arm tracer is separated - see [4], [13]. The Gas flow and old stars, orbiting around a circular orbit around the Galactic Center at 233 km/s, come from the right of the spiral arm in this rendering. The dust and shock at the inner arm edge (red zone) helps some of the gas to contract under gravity to form embedded protostars and masers (orange zone), within 0.5 Myrs of the shock front. These protostars evolve to become newly-formed stars and young HII regions (green zone). Old broad diffuse CO gas congregates in regions near the density wave's Potential Minimum (blue zone). The masers and gas inside the spiral arm goes at a relative speed near 81 km/s away from the density wave's pattern speed (shock front and arm pattern speed going at 152 km/s). The sum of these speeds (152 and 81) equals the absolute speed noted earlier (233).*

The relative speed of the density wave (151.7 km/s) and that of the masers relative to that (an extra 81.3 km/s) means that the masers are going above the arm's inner edge by 233/151.7 = 1.53 times faster (Mach 1.5). At this speed, the masers and young stars would cross the spiral arm, from the Potential Minimum and broad diffuse CO down to the shock and dust lane (separated by about 350 pc) in a time of 350 pc/ 81.3 km/s = 4.3 Myrs.

**Time to reach the next spiral arm.** At the Sun's orbit **near 8.15 kpc of galactic radius**, to cover a quarter of a circle at the mean relative speed from the shock front, one needs 0.25x2x3.14x8150 pc =12 802 pc; at the mean relative speed of 81.3 km/s = 157.5 ±10   Myrs; that is the time for the Earth to experience its passage from one spiral arm to the next arm in the rotating frame of the spiral arm.

**Implications for the density wave.** In a relative frame, stars and gas speed away from the inner arm's shock front at about 81 km/s [13], [1],[ 7].

In an absolute frame, stars and gas orbit around the Galactic Center at 233 km/s; so the velocity difference is the linear pattern speed, being 233 – 81 = 152 km/s. This value of 152 km/s is the linear pattern speed of the density wave, which applies at the solar orbit, so the required angular pattern speed is 152/ 8.1 = 18.8 km/s/kpc..

Near a Galactic radius of 8.1 kpc, that angular pattern speed value near 19 km/s/kpc is in the range suggested by various authors, e.g. [20], and implies a galactic co-rotation radius of  233/18.8 = 12.4  ±1 kpc.

To go from one spiral arm to the next (2 x 3.15 x 8.15kpc / 4), at such a relative speed (81 km/s), requires a time period of 12800 pc / 81 km/s =  158 Myrs (in the rotating frame of the spiral arms, orbiting around the Galactic Center). This would also be the mean time between two major extinctions on Earth - a difficult number to get (even with a large error bar). During that time, the arms would have turned 1.87 times  in their orbit around the Galactic Center (152 km/s x158 Myrs). If the Sun's orbit is not strictly circular but an ellipse, this time value may change somewhat by the amount of the non-zero orbital eccentricity.

Error bars for these numerical deductions are large and not easily ascertained, but the numerical values from these recent extinctions here are not far off the numerical values obtained from the Galactic dynamics of the previous section.

A recent statistical analysis of Earth extinctions obtained a period near 176-188 Myrs, arguably due to successive passages of the Sun and Earth through a Galactic spiral arm ([21]; [22],[23]). Others may differ  (± 20 Myrs) -  successive transits of the Earth through successive spiral arms may have seed crust production every 170-200 Myrs [24] through the Oort's Cloud around the solar system being perturbed by nearby stars (shooting icy constituents down toward the Sun)..

### 4.    Locations of spiral arms – complementarity of arm tangents with radio masers and optical Gaia parallaxes

The precise determinations of the locations of each spiral arm in our Milky Way galaxy could be done using precise distance measurement (parallax of objects inside spiral arms) and also by finding the location in galactic longitude of the tangents from the Sun to tracers inside these spiral arms). These two methods should yield the _**same**_ results for the locations of spiral arms, being complementary in essence.

Catalogs of precise parallax measurements of radio masers with precise distances have been published, as well as a map of arm locations (Fig. 1 in [25]).

Catalogs of precise arm tangents in Galactic longitudes have been published and such arm locations inferred (Vallée [1] with 205 tangents; Vallée [2] with 107 tangents). They led to drawing the map of arm location (Fig.1 in [1] for nearby arms; [14] for distant arms). A comparison between the locations of observed arm tangents and a fitted model of the locations of arm tangents showed a very good fit, as well as a comparison of the arm models from the parallax of radio masers and the arm models model from the arm tangents (Fig.3 in [7]).

The new Gaia DR3 map (Fig. 14 in [5])  shows young open star clusters at optical wavelengths near spiral arms, within 4 kpc of the Sun's location, with the arm locations copied from the locations of radio masers (from Fig.1 in [25]).  But these DR3 young open star clusters are not well aligned with radio masers: thus (1) no new arm fit was done; (2) their distribution is not continuous as there are observed gaps and discontinuities along an arm, (3) there appear to be different optical widths across a radio arm; (4) the young open cluster stars in front of the radio masers are not expected there in some theories.

There is a good complementarity  among the locations of spiral arms. Thus all around the Sun we get the _**same**_ locations of the spiral arms, either through radio masers (distance) or through arm tangents (longitudes).  In addition, the Gaia DR3 optical parallaxes of young star clusters give the _**same**_ locations of spiral arms as radio masers, albeit with a much larger arm width (larger distance errors).

The question of the possible bending of the Perseus arm is mentioned (Section 4 in [26] suggested 2 arms; in contrario, [27] suggested an interarm island near the  Perseus arm). Also,the width of each spiral arm is thus not well defined in Gaia DR3; in contrario, a new multi-tracer approach for defining the spiral arm width was advocated (masers near the Shock front/inner arm  versus broad  diffuse broad CO gas near the Potential Minimum/outer arm – see [28]).

Close to the Galactic Center, each model must start the spiral arms. Our tangent model starts each spiral arm near 2.2 kpc away from the Galactic Center (see Section 1 and Figure 1), and this start value may differ from the parallax models (having larger errors with larger distances).

### 5. Conclusion

We employed the arm model found recently, as fitted to the arm tangents in galactic longitudes, with 8.1 kpc as the Sun to Galactic Center distance (Figure 1).

We made an examination of the observed radial velocity of some arm tracers, at the tangent from the Sun to the spiral arm, yielding the folllowing. Comparing the radial velocity of the Masers at the tangent points, to the same from the broad diffuse CO gas near the Potential Minimum, the prediction of the density wave theory with shocks seems to be valid (Figure 2; Table 1).

A similar comparison, this time taking the [CII] observations near the arm tangent (on-site, and slightly off-site) does not change the statistical results.

Our modeling (Figures 1 and 2) can be employed with current estimates of the passage of Earth through a spiral arm. The ensuing implications are computed for the Galactic angular spiral arm speed (near 22 km/s/kpc) and for the Galactic co-rotation radius (nearer 11 – 12 kpc). The time to reach the next spiral arm is near 158 Myrs. (Section 3).

**Acknowledgements.** The figure production used the PGPLOT software at the NRC Canada in Victoria.

**Data Availability.** All data underlying this article are available in the article (references given below), and / or will be shared on reasonable request to the Corresponding author.

**Funding.** No funds or grants were received during the preparation of this paper. I used the facilities at NRC HAA DAO.

**Declarations: Competing interests**. None.

## References

[1] Vallée. J.P. 2022a. Catalog of spiral arm tangents (Galactic longitudes) in the Milky Way, and the age gradient based on various arm tracers New Astron, 97, 101896 (1-14).

[2] Vallée, J.P. 2014b. Catalog of observed tangents to the spiral arms in the MIlky Wsy galaxy. ApJ Suppl. Ser., 215, 1 (1-9).

[3] Vallée, J.P. 2016a, A substructure inside spiral arms, and a mirror image across the galactic meridian. Astrophys J., vol.821, art.53, pp.1-12.

[4] Vallée, J.P. 2014a. The spiral arms of the Milky Way: the relative location of each different arm tracer, within a typical spiral arm width. Astron J., vol. 148, art.5, pp.1-9.

[5] Vallée, J.P. 2008, New velocimetry & revised cartography of the spiral arms in the Milky Way – a consistent symbiosis. Astron.J., v135, p1301-1310.

[6] Vallée, J.P. 2017a. A guided map to the spiral arms in the galactic disk of the Milky Way. Astronomical Review, vol.13, pp.113-146.

[7] Vallée, J.P. 2022b. The observed age gradient in the Milky Way – as a test of spiral arm structure. Ap Space Sci., 367, 26 (1-10).

[8] Abuter, R., Amorim, A. et al 2019. A geometric distance measurement to the Galactic Center black hole with 0.3% uncertainty. A & A , 625, L10.

[9] Vallée, J.P. 2017b. The Norma spiral arm: large-scale pitch angle. Ap Sp. Sci., 362, 173 (5pp).

[10] Vallée, J.P. 2015. Different studies of the global pitch angle of the Milky Way's spiral arms. MNRAS, 450, 4277-4284.

[11] Vallée, J.P. 2016b. The start of the Sagittarius spiral arm (Sagittarius origin) and the start of the Norma spiral arm (Norma origin): model-computed and observed arm tangents at Galactic longitudes -20º <l< +23º. AJ, 151, 55 (16pp).

[12] Drimmel, R., Poggio, E. 2018. On the Solar velocity. Research Notes A.A.S., 2, art.10.

[13] Vallée, J.P. 2021b. Arm tangents and the spiral structure of the Milky Way –the Age gradient. International J. Astron & Astrophys., 11, 445-457.

[14] Vallée, J.P. 2022c. Superposing the magnetic spiral structure of the Milky Way, on the stellar spiral arms –matching the unique Galactic magnetic field reversal zone with two Galactic spiral arm segments. . International J. Astron & Astrophys., 12, 281-300.

[15] Roberts, W.W. 1969. Large-scale shock formation in spiral galaxies and its implications on star formation. Astrophys. J., 158, 123-143.

[16] Roberts, W.W. 1975. Theoretical aspects of galactic research. Vistas in Astron., 19, 91-109.

[17] Velusamy, T., Langer, W., Pineda, J., Goldsmith,P. 2012. [CII] 158µm line detection of the warm ionized medium in the Scutum-Crux spiral arm tangency. A&A, 541, L10-L13.

[18] Velusamy, T., Langer, W.D., Goldsmith, P.F., Pineda, J.L. 2015, "Internal structure of spiral arms traced with [CII]: unraveling the WIM, HI, and molecular emission lanes", A & A, 578, 135 (1-12).

[19] Vallée, J.P. 2019.  Spatial and velocity offsets of Galactic masers from the centres of spiral arms. MNRAS,489, 2819-2829.
[20] Vallée, J.P. 2021a,  A low density wave's spiral pattern speed, from the tracer separations (age gradient) across a spiral arm in the Milky Way.   MNRAS, 506, 523-530.
[21] Gillman, M. 2022. The galactic signal of impact ages and the ultimate causes of the largest extinctions. In press.
[22] Gillman, M.P., Erenier, H.E., Sutton, P.J. 2019, Mapping the location of terrestrial impacts and extinctions onto the spiral arm structure of the Milky Way.  Internat. J. Astrobiology, 18, 323-328.
[23] Gillman, M., Erenier, H. 2008. The Galactic cycle of extinction. .  Internat. J. Astrobiology, 7, 17-26.
[24] Kirkland, C., Sutton, P., Johnson, T., et al. 2022. Did transit through the Galactic spiral arms seed crust production on the early Earth ?. Geology,  50, 11, 1312-1317.
[25] Reid, M.J., Menten, K.M., Brunthaler, A., et al. 2019. Trigonometric parallaxes of high-mass star-forming regions: our view of  the Milky Way.  ApJ, 885, 131 (1-18).
[26] Drimmel, R.,  Romero-Gomez, M., Chemin, L., et al 2022. Gaia Data Release 3 : mapping the asymmetric disc of the Milky Way. A&A, in press. arXiv:2206.06207 astro-ph.GA
[27] Vallée, J.P. 2020a. Interarm islands in the Milky Way – the one near the Cygnus spiral arm. MNRAS, 494, 1134-1142.
[28] Vallée, J.P.  2020b. A new multitracer approach to defining the spiral arm width in the Milky Way. Ap J 896, 19 (1-10).